\documentclass{PoS}
\usepackage{subfig}
\usepackage{epsfig,graphicx}
\usepackage[pdf]{pstricks}
\usepackage[crop=off]{auto-pst-pdf}
\usepackage{amsmath}
\usepackage{changepage}
\usepackage{epstopdf}
\newcommand{\beq}{\begin{equation}}
\newcommand{\eeq}{\end{equation}}

\def\refeq#1{\mbox{(\ref{#1})}}

\def\reffi#1{\mbox{Figure~\ref{#1}}}

\def\refta#1{\mbox{Table~\ref{#1}}}

\def\citere#1{\mbox{Ref.~\cite{#1}}}

\newcommand{\GeV}{\unskip\,\mathrm{GeV}}

\newcommand{\fb}{\unskip\,\mathrm{fb}}

\def\mathswitchr#1{\relax\ifmmode{\mathrm{#1}}\else$\mathrm{#1}$\fi}

\newcommand{\PW}{\mathswitchr W}

\newcommand{\Pne}{\mathswitch \nu_{\mathrm{e}}}

\newcommand{\Pp}{\mathswitchr p}

\newcommand{\Pep}{\mathswitchr {e^+}}

\newcommand{\Pmup}{\mathswitchr {\mu^+}}

\newcommand{\PWp}{\mathswitchr {W^+}}

\newcommand{\jet}{\mathswitchr {j}}

\def\mathswitch#1{\relax\ifmmode#1\else$#1$\fi}

\newcommand{\f}{\frac}
\newcommand{\nn}{\nonumber}

\newcommand{\re}{\mathrm{e}}
\newcommand{\rj}{\mathrm{j}}
\newcommand{\rd}{\mathrm{d}}

\newcommand{\rp}{\mathrm{p}}

\newcommand{\muR}{\mu_{\mathrm R}}
\newcommand{\muF}{\mu_{\mathrm F}}
\newcommand{\pT}{p_{\mathrm T}}
\newcommand{\rT}{\mathrm T}

\title{NLO QCD corrections to $\Pp\Pp\to
\Pep\Pne\mu^+\nu_\mu \jet\jet$ in vector-boson fusion at the LHC}

\ShortTitle{NLO QCD corrections to $\Pp\Pp\to
\Pep\Pne\mu^+\nu_\mu \jet\jet+X$ in VBF at the LHC}

\author{\speaker{Lucia HO\v SEKOV\' A}\\
         \thanks{LPN 13-075, IFIC 13-74}\\
        Instituto de F\'isica Corpuscular, Universitat de Val\`encia - Consejo Superior de Investigaciones Cient\'ificas,\\
        Parc Cient\'ific, 46980 Paterna (Valencia), Spain\\
        E-mail: \email{lucia.hosekova@ific.uv.es}}

\abstract{
We report on studies of next-to-leading order (NLO) QCD corrections to the process \mbox{$\Pp\Pp\rightarrow\Pep\Pne\Pmup{\nu}_\mu \jet\jet$} which is
associated with vector-boson fusion (VBF) production of an intermediate $\PW^+\PW^+$ pair. The impact of $s$-channel
diagrams and interference contributions in VBF kinematics is analyzed at leading order (LO) and found to be entirely negligible.  The NLO corrections are around 5\% of the
LO cross section, and the scale dependence is reduced to about ~$1\%$. We introduce a dynamic scale which improves the $K$ factor in the
high-energy tails of the distributions.}

\FullConference{The European Physical Society Conference on High Energy Physics \\
         18-24 July, 2013\\
         Stockholm, Sweden}

\begin{document}

\section{Introduction}

VBF processes at the Large Hadron Collider (LHC), due to their
unique signature formed by two tagging jets in the forward and
backward region of the detector, offer interesting possibilities
for discovery and exploring properties of the Higgs boson,
testing electroweak symmetry breaking and for various searches for
new physics.

We focus on the process involving the scattering of two positively
charged W bosons via VBF with subsequent leptonic decay, leading
to a final state with two jets, two positively charged leptons and
two neutrinos at LO, i.e. \mbox{$\Pp\Pp\to \PWp\PWp \jet\jet+X \to
\Pep\Pne\mu^+\nu_\mu \jet\jet+X$} with a distinct signature
of same-sign high-$\pT$ leptons, missing energy and extra jets. We are
specifically interested in the electroweak (EW) production mode of the order
$\alpha_{\rm{EW}}^6$ which includes genuine VBF diagrams where the
scattering weak gauge bosons are emitted from the incoming
(anti-)quarks. We include both resonant
contributions where the final-state leptons are produced via
vector-boson decay as well as non-resonant ones.

The work presented in this report represents an independent calculation
and verification of \citere{Jager:2009xx}, and includes an assessment of the $s$-channel and interference
contributions at LO.
More details on the computational aspects of the calculation, and a comprehensive discussion of the numerical results
and differential distributions of kinematical observables, can be found in \citere{Denner:2012dz}.

\section{Technical aspects}
From computational standpoint, the reactions of type \mbox{$\rp
\rp \to 4l\rj\rj+X$} pose a challenging problem due to large
number of contributing diagrams. Our strategy for dealing with the
increased complexity lies in creating a database of building
blocks and taking advantage of the fact that the EW and QCD parts of
the diagrams are largely independent of one another.

Due to the fact that LHC experiments are conducted at TeV energies, fermion mass effects are
strongly suppressed and have been neglected in this calculation. Furthermore, the contributions from the third-generation quarks and leptons
is expected to be small \cite{Denner:2012dz} and have also been neglected, which allows us to approximate the CKM matrix by a unit matrix
provided the interferences between kinematic channels as well as the $s$-channel contributions are negligible  \cite{Bozzi:2007ur}.

In order to create the building block database, all contributing diagrams are factorized by inserting the polarization
sums for off-shell massive gauge bosons
\beq\label{polsum}
g^{\mu\nu}=-\sum_{i=\{+,-,0\}}\varepsilon^\mu_i(k)\varepsilon^{*\nu}_i(k)+\f{k^\mu k^\nu}{k^2}\;
\eeq
in place of internal vector bosons connecting the EW and QCD sections of the diagrams (gauge-boson mass in the definition of longitudinal polarization vectors is replaced by $\sqrt{k^2}$).
This method allows us to reuse the building blocks in multiple instances, in particular the EW blocks which are not effected by QCD corrections and can be quite complicated by themselves.

\begin{figure}
\begin{adjustwidth}{-2em}{-2em}
\centering
\includegraphics[bb = 100 520 595 772, height=5cm]{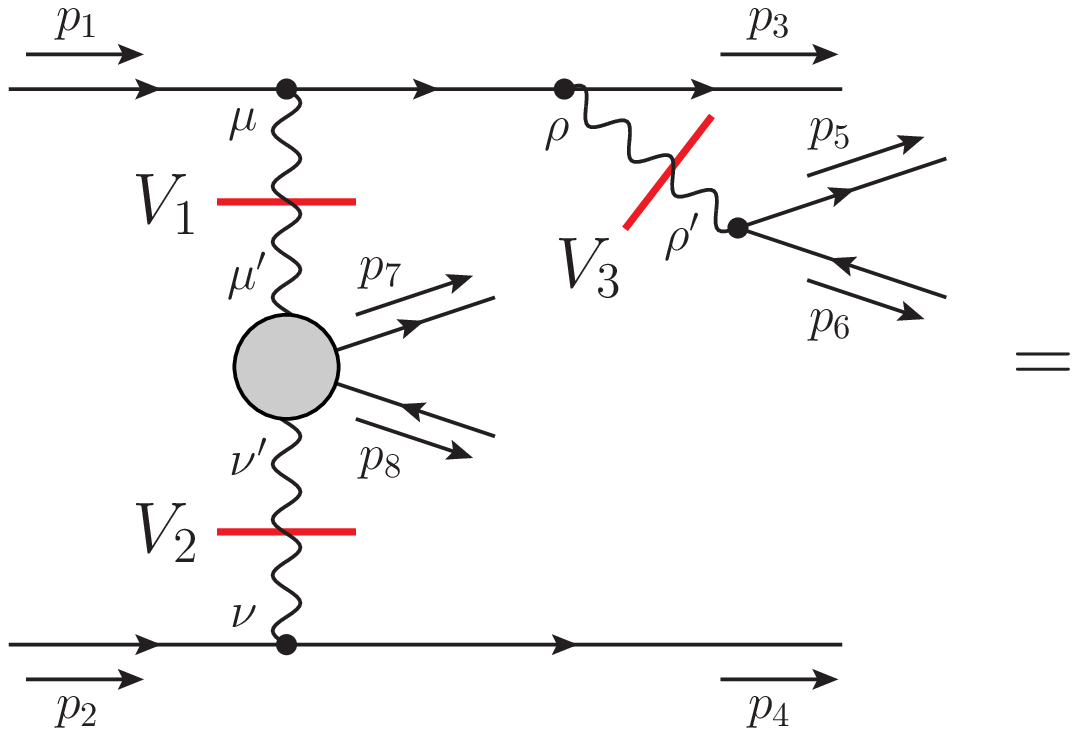}
\includegraphics[bb = 140 480 500 772, height=4.8cm]{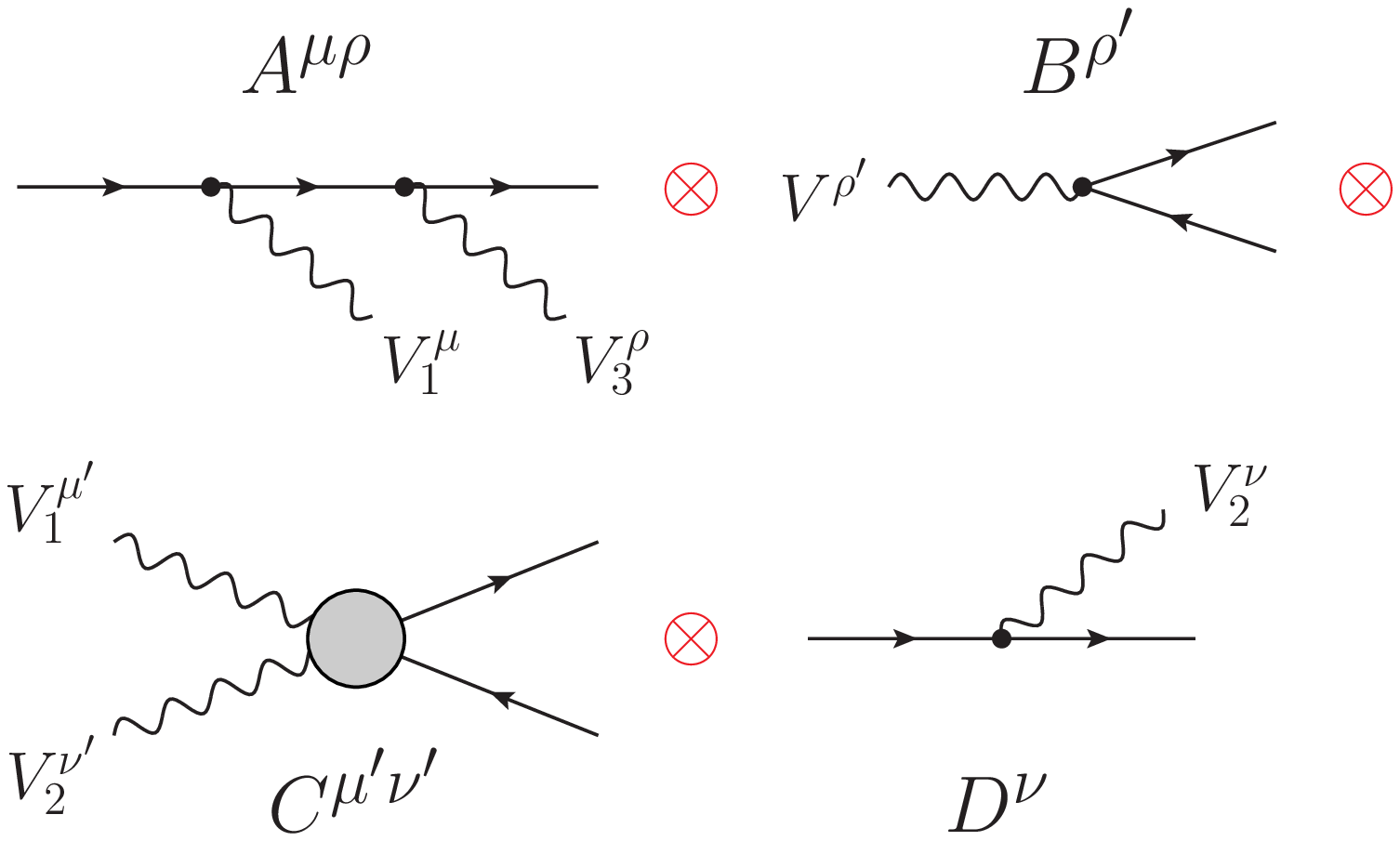}
\centering
\end{adjustwidth}
\caption{ Example of a LO diagram split into four building blocks by applying the polarization sums
  (\protect\ref{polsum}) to cut three intermediate vector bosons.}
\label{polsum2}
\end{figure}

\reffi{polsum2} shows an example of a contributing diagram at LO split into four building blocks by inserting three polarization sums. The resulting amplitude can be written as
\begin{align}\label{eq:ampsplit}
-\sum_{i,j,k={\pm,0,m}}& [A^{\mu\rho}\varepsilon^*_{i,\mu}(k_1)\varepsilon^*_{k,\rho}(k_3)]\;[B^{\rho'}\varepsilon_{k,\rho'}(k_3)]\;[C^{\mu'\nu'}\varepsilon_{i,\mu'}(k_1)
\varepsilon_{j,\nu'}(k_2)]\;[D^\nu\varepsilon^*_{j,\nu}(k_2)]\nn\\
&\times \frac{1}{k_1^2-M_{V_1}^2}  \frac{1}{k_2^2-M_{V_2}^2}\frac{1}{k_3^2-M_{V_3}^2},
\end{align}
where we have denoted $\varepsilon_m^\mu(k)=\f{k^\mu}{\sqrt{k^2}}$ and $\varepsilon_m^{*\mu}(k)=-\f{k^\mu}{\sqrt{k^2}}$ in order to compactify the expression. \mbox{$1/{(k_i^2-M_{V_i}^2)}$}
are denominator parts of the split gauge boson propagators.

The expressions for individual building blocks are obtained using the $\textsc{FormCalc}$ 6 package~\cite{Hahn:2009bf}, modified to accommodate the form \refeq{eq:ampsplit} and exported into
\textsc{Fortran} modules, while the formulas for combining the blocks are implemented as \textsc{Fortran} subroutines. At NLO, the virtual corrections lead to tensor integrals given by  two- to five-point functions. Tensor reduction is performed numerically in \textsc{Fortran} by the means of the  \textsc{Collier}
library~\cite{Denner:2002ii,Denner:2005nn,Denner:1991qq,Beenakker:1988jr,Denner:2010tr,Coli6}. Infrared divergencies are controlled using the Catani--Seymour dipole-subtraction technique
for massless particles~\cite{Catani:1996vz}, while the  ultraviolet divergencies are renormalized
by adding corresponding counterterms. The \textsc{Fortran}
code for each process is contained in a single function which is
called from within a Monte Carlo program originally developed for the calculation published in \citere{Denner:2012yc}. For practical reasons, we
used the tree-level amplitudes generated with \textsc{OpenLoops}
\cite{Cascioli:2011va} after cross-checking them against the ones
obtained in the approach described above.  The virtual amplitudes,
on the other hand, are constructed using the building-block method.

Correctness of the calculation has been verified by comparing with available results and tools at both LO and NLO. Matrix elements have been compared with \textsc{MadGraph~4}~\cite{Alwall:2007st}, \textsc{OpenLoops}~\cite{Cascioli:2011va} and stand-alone $\textsc{FormCalc}$ 6~\cite{Hahn:2009bf} combined with \textsc{LoopTools} for a set of phase-space points.

We have reproduced the calculation of the full integrated cross section for the NLO QCD corrections to \mbox{$\rp \rp\to  \re^+\nu_{\re}\mu^+\nu_{\mu}\rj\rj+X$} published in
\citere{Jager:2009xx} with the same setup and input parameters.  Assuming a statistical error of the results of \citere{Jager:2009xx} of per-mille (it is stated to be at the
sub-per-mille level), the difference amounts to only $2\sigma$ and is thus acceptable. More detailed discussion of this comparison can be found in Ref. \cite{Denner:2012dz}.

\section{Numerical results}

\begin{figure}
\centering
\subfloat[$\muF=\muR=\xi M_{\mathrm W}$]
{\label{FS}\epsfig{figure=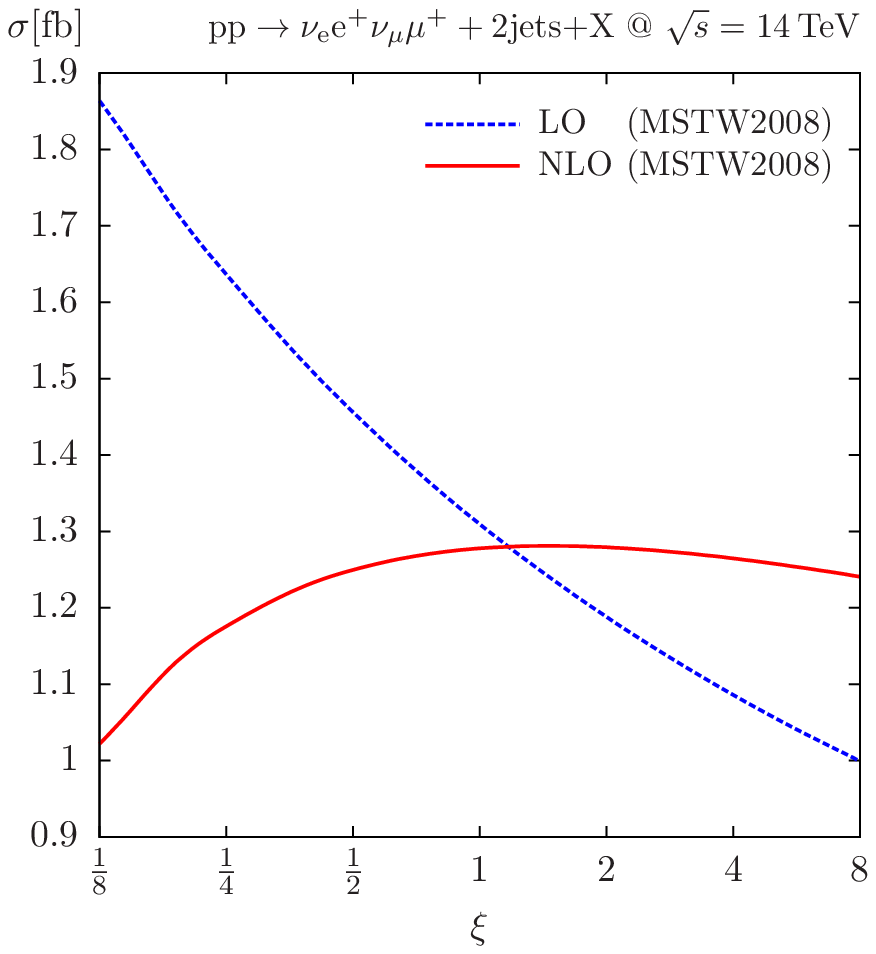,width=7cm}}\qquad
\subfloat[$\muF=\muR=\xi \sqrt{p_{\rm T,\mathrm{jet}_1}\cdot p_{\rm T,\mathrm{jet}_2}}$]
{\label{DS2}\epsfig{figure=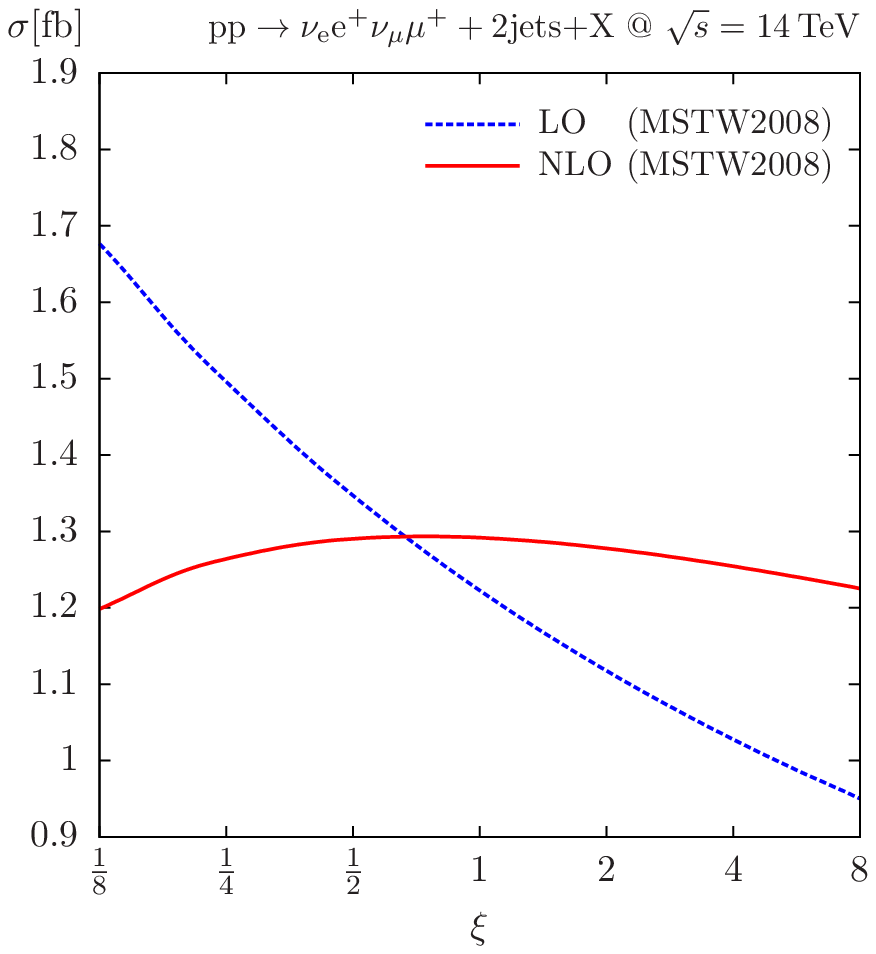,width=7cm}}
\caption{Scale dependence of the LO (dotted blue line) and NLO (solid red line) cross section for the fixed (a) and
dynamic scale (b) as a function of the scale parameter $\xi$.}
\label{scales}
\end{figure}

In order to enhance regions of the phase-space where VBF-type processes can be observed
experimentally, we impose a number of kinematic cuts at the Monte Carlo level. Following the proposal in \citere{Jager:2009xx}, these include two hard jets with large
rapidity separation and charged final-state leptons located in the central detector region
and separated from the jet activity. The complete set of cuts is listed in \citere{Denner:2012dz}.

We have chosen two types of scales to demonstrate their effects on
the behaviour of the NLO distributions for selected observables.
In the fixed-scale (FS) choice, both factorization and
renormalization scales have been set to the mass of the W boson
and varied by a factor $\xi$ around this central value,
\beq\label{FS1}
\muF=\muR=\xi M_{\mathrm W}.
\eeq
Since this FS choice turns out to result in strongly phase-space dependent $K$ factors---%
in particular in the high-energy tails of distributions---a dynamical
scale (DS),
\beq\label{DS}
\muF=\muR=\xi \sqrt{p_{\mathrm{T,j_1}}\cdot p_{\mathrm{T,j_2}}}.
\eeq
has been considered as well. The dependence of the total cross section on the parameter $\xi$ for
both scale choices is depicted in \reffi{scales}.  The scale variation of the LO cross section
which only depends on $\xi$ via $\muF$ is about $\pm10\%$,
while at NLO it is reduced to about $\pm2\%$ of the total cross
section for the FS choice and $\pm1\%$ for the DS choice.

\begin{figure}[htb]
  \centering \subfloat[$\muF=\muR=M_{\mathrm W}$]
  {\label{pTmaxFS}\epsfig{figure=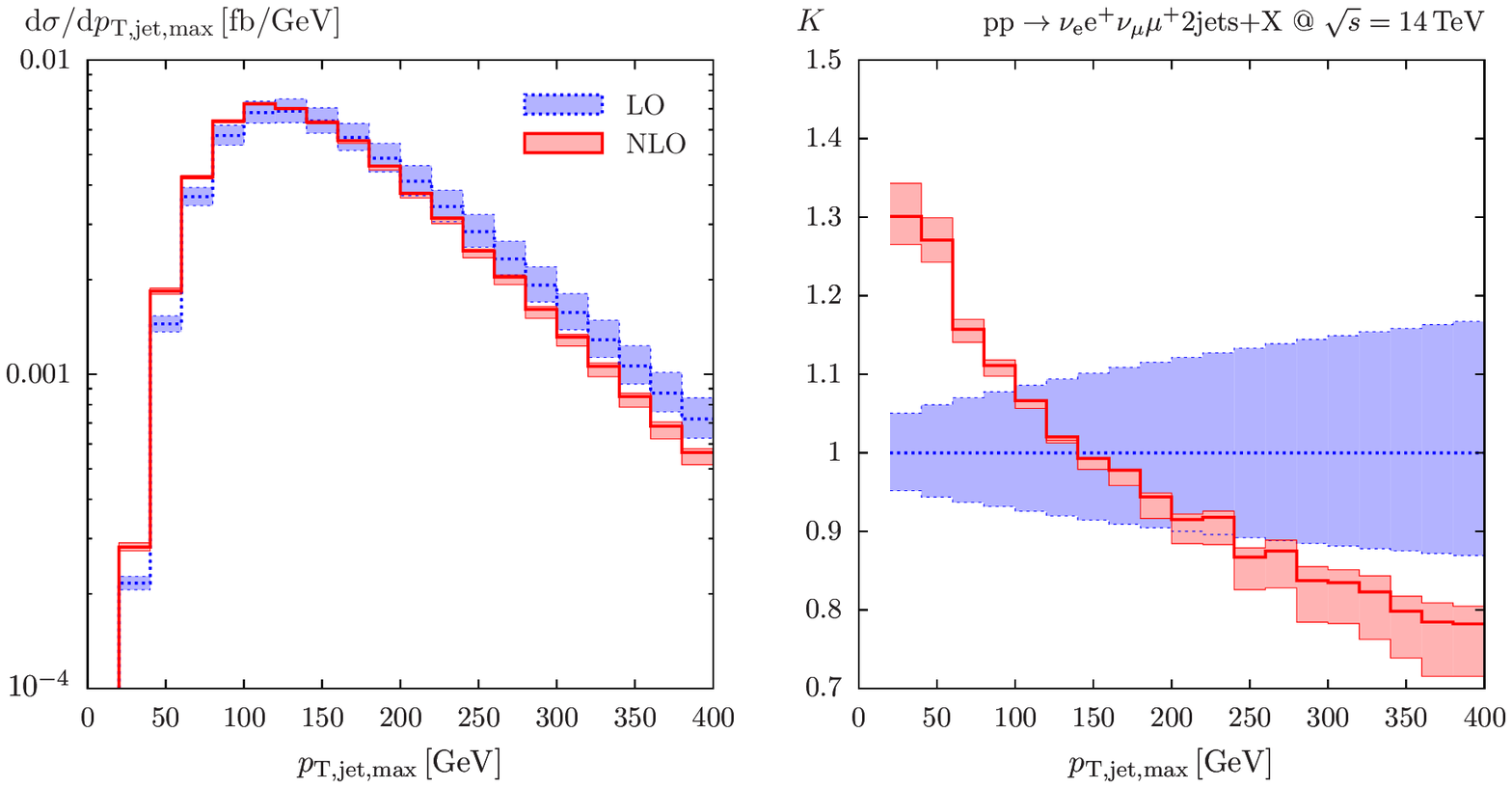,width=14cm}}\\
  \subfloat[$\muF=\muR=\sqrt{p_{\rm T,\mathrm{jet}_1}\cdot p_{\rm
      T,\mathrm{jet}_2}}$]
  {\label{pTmaxDS2}\epsfig{figure=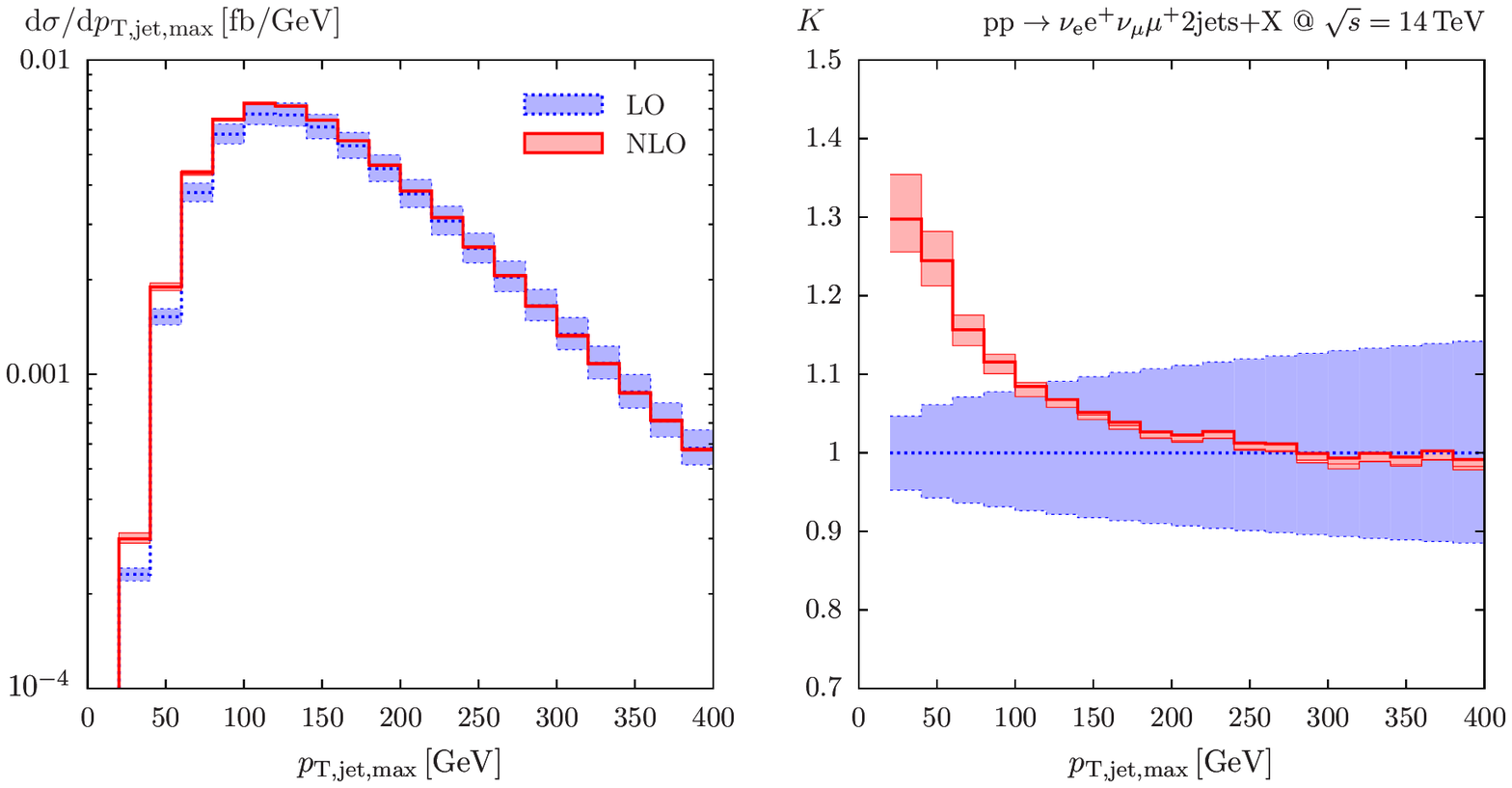,width=14cm}}
\caption{Transverse momentum distribution of the tagging jet with the higher $\pT$ for the fixed (a) and
  dynamic scale (b) on the left and the
  corresponding $K$ factor represented by the solid (red) line on the
  right.}
\label{pTmax plot}
\end{figure}

\begin{figure}[htb]
\centering
\epsfig{figure=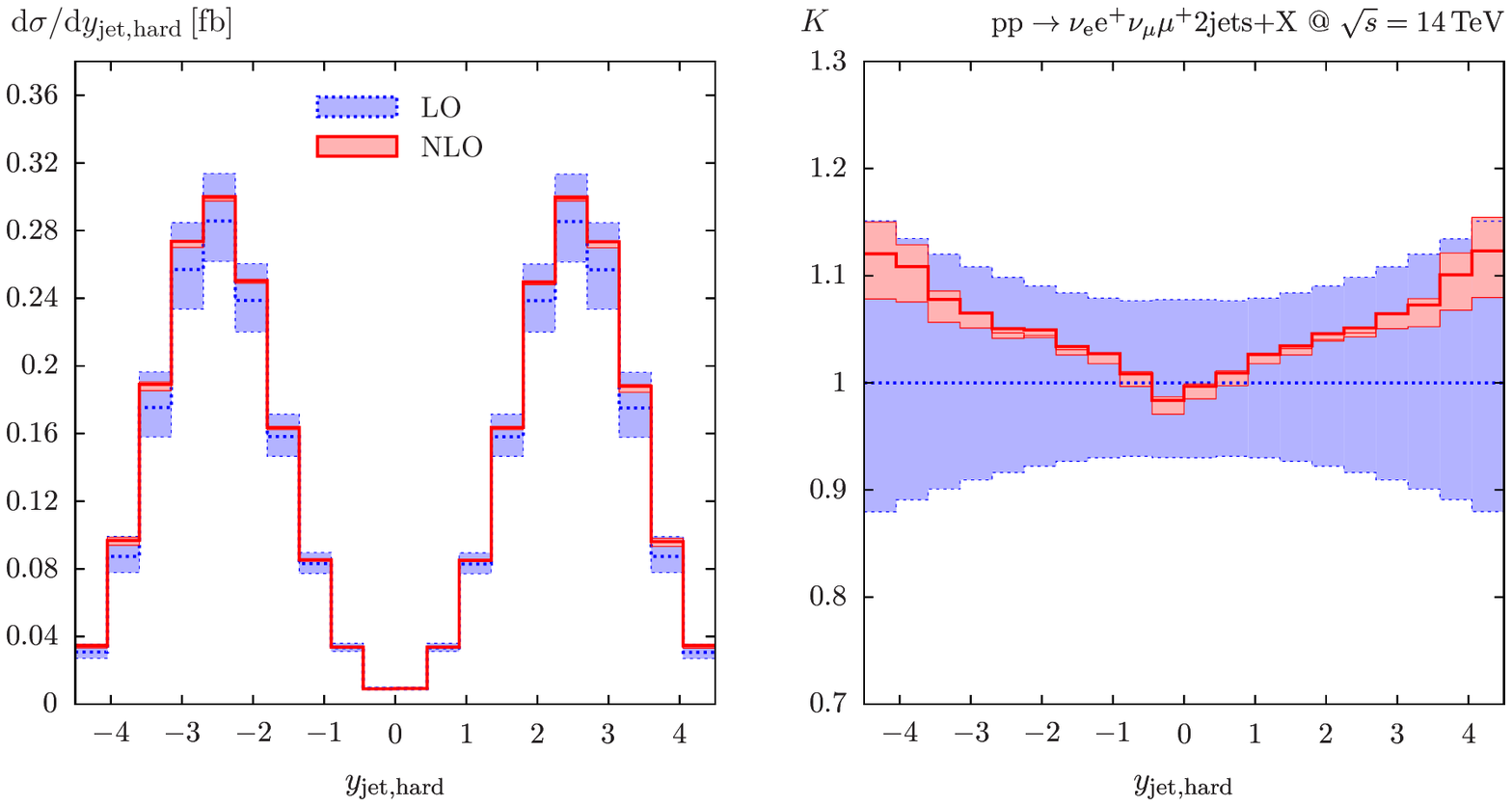,width=14cm}
\caption{Absolute rapidity distribution for the harder tagging jet  for the
dynamic scale on the left and the corresponding $K$ factor represented by the solid (red) line on the right.}
\label{absyhard plot}
\end{figure}

Dedicated VBF cuts prefer $t$- and $u$-channel kinematics, while the $s$-channel configurations as well as interferences between $t$ and $u$ channels are strongly suppressed.
It can be argued~\cite{Jager:2009xx} that these contributions can be safely neglected. In order to demonstrate this claim, the LO cross section has been evaluated
for three different sets of matrix elements as shown in \refta{finalcs}. It confirms that $\sigma^{\mathrm{LO}}_{\mathrm{VBF}}$ (containing only squares of $t$- and $u$-channel contributions but no
interferences) can
be considered a very good approximation of the full LO cross section, and the NLO cross section has therefore been evaluated using only
$t$- and $u$-channel contributions without interferences between them
in order to improve the speed of the calculation.
\begin{table}
  \renewcommand*\arraystretch{1.3}
  \centering
  \begin{tabular}{ccc|c}
    \hline
    $\sigma^{\mathrm{LO}}_{\mathrm{full}}[\fb]$ & $\sigma^{\mathrm{LO}}_{\mathrm{VBF+int}}[\fb]$ & $\sigma^{\mathrm{LO}}_{\mathrm{VBF}}[\fb]$ & $\sigma^{\mathrm{NLO}}_{\mathrm{VBF}}[\fb]$ \\
    \hline
     1.2224(2) &       1.2218(2) &       1.2230(2) &    1.2917(8) \\
    \hline
  \end{tabular}
\vspace*{3ex}
  \caption{Integrated cross sections for LO including all channels and interferences ($\sigma^{\mathrm{LO}}_{\mathrm{full}}$),
    for LO including $t$--$u$ interferences, but neglecting $s$-channel diagrams ($\sigma^{\mathrm{LO}}_{\mathrm{VBF+int}}$),
    for LO in the VBF approximation, i.e.\ neglecting all $s$-channel diagrams and interferences ($\sigma^{\mathrm{LO}}_{\mathrm{VBF}}$), and
    for NLO in the VBF approximation
    ($\sigma^{\mathrm{NLO}}_{\mathrm{VBF}}$).}
  \label{finalcs}
\end{table}

\subsection{Differential distributions}

\reffi{pTmax plot} shows the LO and NLO cross sections as a function of
the transverse momentum of the harder of the two
tagging jets. The distribution
peaks at \mbox{$p_{\rT} \sim110\GeV$}, confirming the preference of the high-$\pT$
regions by the VBF tagging jets. The plots on the right show the LO
and NLO predictions normalized to the LO result at the central scale,
i.e.\ \mbox{$K_{\mathrm{LO}}(\xi)=\rd \sigma_{\mathrm{LO}}(\xi)/\rd\sigma_{\mathrm{LO}}(\xi=1)$}
(dotted blue line), and
\mbox{$K_{\mathrm{NLO}}(\xi)=\rd \sigma_{\mathrm{NLO}}(\xi)/\rd\sigma_{\mathrm{LO}}(\xi=1)$}
(solid red line).
One can observe that $K(p_{\mathrm{T,j_{max}}})$ grows noticeably in
low-$\pT$ regions for both FS and DS towards the value 1.3, while in
the larger-$\pT$ regions it drops to 0.8 in case of the FS
(\reffi{pTmaxFS}) and remains very close to 1 for the DS
(\reffi{pTmaxDS2}), which is a behaviour that motivated the choice of the
DS in the first place and can also be observed in other jet distributions.

Rapidity of the tagging jets is another distinguishing feature of the VBF processes since
they exhibit very little jet activity in its central region, as shown in \reffi{absyhard plot}. This is in sharp contrast to the
behaviour of the QCD production mode for  $\PWp\PWp$, where the jet rapidity peaks
at 0, dominating the central rapidity region~\cite{Jager:2011ms}.  This process thus constitutes background which can be suppressed dramatically by imposing a cut on the separation of individual jet rapidities $\Delta
y_{\mathrm{\rj\rj}}$.

\acknowledgments
I would like to thank Ansgar Denner and Stefan Kallweit for their fruitful collaboration, and the theory groups at PSI and the
University of Zurich for their hospitality. This work is supported in part by the European Commission through the
Marie-Curie Research Training Network HEPTOOLS under contract
MRTN-CT-2006-035505, by the Research Executive Agency (REA)
of the European Union under the Grant Agreement number PITN-GA-2010-264564 (LHCPhenoNet), by the Spanish Government and EU ERDF funds (grants FPA2007-60323, FPA2011-23778 and CSD2007-00042 Consolider Project CPAN)
and by GV (PROMETEUII/2013/007).

\bibliography{proceedings_arXiv}

\providecommand{\href}[2]{#2}\begingroup\raggedright\begin{thebibliography}{10}

\bibitem{Jager:2009xx}
B.~{J\"ager}, C.~Oleari, and D.~Zeppenfeld, {\it {Next-to-leading order QCD
  corrections to $W^+W^+jj$ and $W^-W^-jj$ production via weak-boson fusion}},
  {\em Phys. Rev.} {\bf D80} (2009) 034022,
  [\href{http://xxx.lanl.gov/abs/0907.0580}{{\tt arXiv:0907.0580}}].

\bibitem{Denner:2012dz}
A.~Denner, L.~Ho{\v s}ekov{\'a}, and S.~Kallweit, {\it {NLO QCD corrections to
  W+ W+ jj production in vector-boson fusion at the LHC}},  {\em Phys.Rev.}
  {\bf D86} (2012) 114014, [\href{http://xxx.lanl.gov/abs/1209.2389}{{\tt
  arXiv:1209.2389}}].

\bibitem{Bozzi:2007ur}
G.~Bozzi, B.~{J\"ager}, C.~Oleari, and D.~Zeppenfeld, {\it {Next-to-leading
  order QCD corrections to $W^+Z$ and $W^-Z$ production via vector-boson
  fusion}},  {\em Phys. Rev.} {\bf D75} (2007) 073004,
  [\href{http://xxx.lanl.gov/abs/hep-ph/0701105}{{\tt hep-ph/0701105}}].

\bibitem{Hahn:2009bf}
T.~Hahn, {\it {FormCalc 6}},  {\em PoS} {\bf ACAT08} (2008) 121,
  [\href{http://xxx.lanl.gov/abs/0901.1528}{{\tt arXiv:0901.1528}}].

\bibitem{Denner:2002ii}
A.~Denner and S.~Dittmaier, {\it {Reduction of one-loop tensor 5-point
  integrals}},  {\em Nucl. Phys.} {\bf B658} (2003) 175--202,
  [\href{http://xxx.lanl.gov/abs/hep-ph/0212259}{{\tt hep-ph/0212259}}].

\bibitem{Denner:2005nn}
A.~Denner and S.~Dittmaier, {\it {Reduction schemes for one-loop tensor
  integrals}},  {\em Nucl. Phys.} {\bf B734} (2006) 62--115,
  [\href{http://xxx.lanl.gov/abs/hep-ph/0509141}{{\tt hep-ph/0509141}}].

\bibitem{Denner:1991qq}
A.~Denner, U.~Nierste, and R.~Scharf, {\it {A Compact expression for the scalar
  one loop four point function}},  {\em Nucl. Phys.} {\bf B367} (1991)
  637--656.

\bibitem{Beenakker:1988jr}
W.~Beenakker and A.~Denner, {\it {Infrared Divergent Scalar Box Integrals with
  Applications in the Electroweak Standard Model}},  {\em Nucl. Phys.} {\bf
  B338} (1990) 349--370.

\bibitem{Denner:2010tr}
A.~Denner and S.~Dittmaier, {\it {Scalar one-loop 4-point integrals}},  {\em
  Nucl.Phys.} {\bf B844} (2011) 199--242,
  [\href{http://xxx.lanl.gov/abs/1005.2076}{{\tt arXiv:1005.2076}}].

\bibitem{Coli6}
A.~Denner, S.~Dittmaier, and L.~Hofer, ``{\sl COLLIER, Complex One-Loop Library
  In Extended Regularizations}.'' in preparation.

\bibitem{Catani:1996vz}
S.~Catani and M.~H. Seymour, {\it {A general algorithm for calculating jet
  cross sections in NLO QCD}},  {\em Nucl. Phys.} {\bf B485} (1997) 291--419,
  [\href{http://xxx.lanl.gov/abs/hep-ph/9605323}{{\tt hep-ph/9605323}}].

\bibitem{Denner:2012yc}
A.~Denner, S.~Dittmaier, S.~Kallweit, and S.~Pozzorini, {\it {NLO QCD
  corrections to off-shell top-antitop production with leptonic decays at
  hadron colliders}},  {\em JHEP} {\bf 1210} (2012) 110,
  [\href{http://xxx.lanl.gov/abs/1207.5018}{{\tt arXiv:1207.5018}}].

\bibitem{Cascioli:2011va}
F.~Cascioli, P.~Maierhofer, and S.~Pozzorini, {\it {Scattering Amplitudes with
  Open Loops}},  {\em Phys.Rev.Lett.} {\bf 108} (2012) 111601,
  [\href{http://xxx.lanl.gov/abs/1111.5206}{{\tt arXiv:1111.5206}}].

\bibitem{Alwall:2007st}
J.~Alwall, P.~Demin, S.~de~Visscher, R.~Frederix, M.~Herquet, et~al., {\it
  {MadGraph/MadEvent v4: The New Web Generation}},  {\em JHEP} {\bf 0709}
  (2007) 028, [\href{http://xxx.lanl.gov/abs/0706.2334}{{\tt
  arXiv:0706.2334}}].

\bibitem{Jager:2011ms}
B.~{J\"ager} and G.~Zanderighi, {\it {NLO corrections to electroweak and QCD
  production of $W^+W^+$ plus two jets in the POWHEG BOX}},  {\em JHEP} {\bf
  1111} (2011) 055, [\href{http://xxx.lanl.gov/abs/1108.0864}{{\tt
  arXiv:1108.0864}}].

\end{thebibliography}\endgroup

\end{document}